# LUMINOSITY OF A COLLIDER WITH ASYMMETRIC BEAMS

*I.N. Meshkov[1]*

Joint Institute for Nuclear Research, Dubna, Russia

Sankt Peterburg University, Sankt Peterburg, Russia

Abstract

A formula for the collider luminosity is derived for a head-on collision of two beams with different parameters. The formula is valid for colliding and partially separated ("merging") beams. Three particular cases are presented: the collision of two identical axially symmetric bunches, the collision of a bunch with a disbanded beam, and the collision of two disbanded bunches. The colliding beams have coinciding longitudinal axes. The formula is valid for colliding both counter propagating and co-propagating beams.

## INTRODUCTION

The problem of calculating the luminosity of a collider is known since the appearance of the first accelerators with colliding beams. Nevertheless, a fairly compact formula describing the luminosity of the collider in the general case of collision of two beams with arbitrary parameters, required to perform analytical or numerical calculations, has not yet been proposed. Such a formula is helpful in selection of collider parameters. The formulas for collisions of two identical bunches are well-known (see, for example, [1 - 3]). Attempts to obtain a formula for a general case lead, as a rule, to cumbersome expressions containing multiple integrals over the six coordinates of 6D phase-space [4].

In this paper we consider the collision of two beams which differ in the parameters – size, emittance, kind of particles and their energy.

## 1. ASYMMETRIC COLLIDING BEAMS - GENERAL CASE

The particle density distribution for each bunch of $N_i$ particles is assumed to be Gaussian in all three dimensions ($x$, $y$, $s$):

$$\rho_i(t) = \frac{N_i}{(2\pi)^{3/2}\,\sigma_{xi}(t)\cdot\sigma_{yi}(t)\cdot\sigma_{si}} \cdot \exp\left\{-\frac{1}{2}\left(\frac{x^2}{\sigma_{xi}^2(t)} + \frac{y^2}{\sigma_{yi}^2(t)} + \frac{s^2}{\sigma_{si}^2}\right)\right\},$$

$$i = 1, 2.$$

In the case of coasting beam of $N_i$ particles, when a uniform density distribution along the circumference is assumed, an analogous formula for the density has the form:

---

[1] E-mail: meshkov@jinr.ru

$$\rho(t) = \frac{N_i}{2\pi\sigma_{xi}(t)\cdot\sigma_{yi}(t)\cdot C_{ring}}\cdot\exp\left\{-\frac{1}{2}\left(\frac{x^2}{\sigma_{xi}^2(t)}+\frac{y^2}{\sigma_{yi}^2(t)}\right)\right\}. \tag{2}$$

In calculations we choose that the centers of both bunches are at the origin of coordinate frame ($x = y = s = 0$) at $t = 0$ (Fig. 1). Then the centers of the bunches $s_i^0$ change with time as

$$s_i^0(t) = v_i t, \quad x_i = y_i = 0 \quad , \tag{3}$$

which implies that the collisions are head-on. As will be shown, the result obtained below is valid for both counter propagating, and co-propagating (or "merging") bunches.

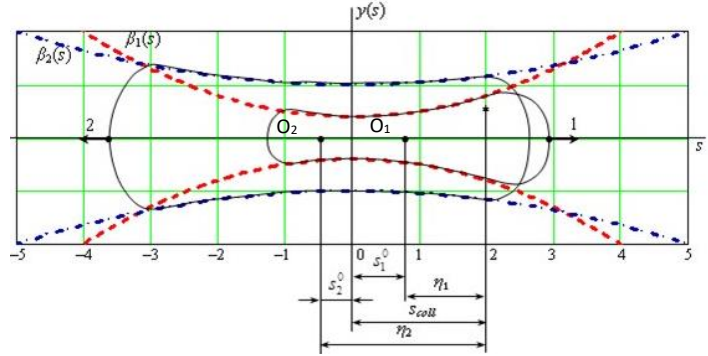

Fig.1. Scheme of collision of two bunches. $O_{1,2}$ - centers of bunches, $\eta_{1,2}$ - distances of the collision point * of two particles from the centers of bunches, $s_{coll}$ - the same, from the origin; $\beta_{1,2}$ - envelopes (beta functions) of the first and second beams.

A dependence of bunch transverse dimensions on time is:

$$\sigma_{\alpha i}(t) = \sqrt{\varepsilon_{\alpha i}\beta_{\alpha i}}, \quad \alpha = x, y, \quad i = 1,2 \quad , \tag{4}$$

where $\varepsilon_{\alpha i}$ is the bunch transverse emittance; and we assume an absence of focusing near the interaction point which results in that:

$$\beta_{\alpha i}(s_i(t)) = \beta_{\alpha i}^* + \frac{s_i^2(t)}{\beta_{\alpha i}^*} \quad . \tag{5}$$

Here $\beta_{\alpha i}^*$ is the minimum value of the beta-function, $s_i(t)$ is the particle coordinate at time $t$.

We also assume the dispersion in the particle interaction section equal to zero (which is usually used in practice).

We now introduce the coordinates of the particles in the systems of moving bunches:

$$x_i', y_i', \eta_i \quad .$$

Obviously, $x_i' = x_i$, $y_i' = y_i$, and the $s$-coordinate of a particle in the laboratory system is equal to:

$$s_i(t) = s_i^0(t) + \eta_i . \tag{6}$$

Accordingly, the $s$-coordinate of the two colliding particles of bunches 1 and 2 is

$$s_{coll}(t) = s_1^0(t) + \eta_1 = s_2^0(t) + \eta_2 \quad . \tag{7}$$

Hence, taking into account conditions (3), we find

$$\eta_2 = \eta_1 + \left(1 - \frac{v_2}{v_1}\right) \cdot s_1^0.$$  (8)

Introducing the designation $\xi \equiv s_1^0, \quad \eta \equiv \eta_1$ we can write down

$$\eta_2 = \eta + V\xi, \quad V \equiv 1 - \frac{v_1}{v_2}.$$  (9)

The Jacobian of the transition is equal to

$$\frac{D(\eta_1, \eta_2)}{D(\xi, \eta)} = V.$$

During the collision, a layer of particles of the first bunch $\rho_1(x, y, \eta_1, t) \cdot d\eta_1$ crosses bunch 2, colliding at each point $s_{coll}$ with a layer $\rho_2(x, y, \eta_2, t) \cdot d\eta_2$ of bunch 2. Since we expressed $\eta_2$ in terms of $\eta$ and $\xi(t) = v_1 t$, integration over the collision time can be replaced by integration over $\xi$ (the coordinate of the center of the first bunch), and integration over the lengths of the bunches can be carried over the variable $\eta$. Now one can write an expression for the luminosity,

$$L = n_{bunch} \cdot f_0 \int_{-\infty}^{\infty} dx \int_{-\infty}^{\infty} dy \int_{-\infty}^{\infty} d\eta_1 \int_{-\infty}^{\infty} d\eta_2 \cdot \rho_1(x, y, \eta_1(t)) \cdot \rho_2(x, y, \eta_2(t))$$  (10)

where $n_{bunch}$ is the number of bunches (see below section 5), and $f_0$ is the particle revolution frequency. For convenience, one can rewrite this expression in the following form:

$$L = \frac{n_{bunch} N_1 N_2 f_0}{(2\pi)^3 \sigma_{s1} \sigma_{s2}} \cdot V \cdot \int_{-\infty}^{\infty} d\xi \int_{-\infty}^{\infty} d\eta \int_{-\infty}^{\infty} \psi_x(x) dx \int_{-\infty}^{\infty} \psi_y(y) dy \cdot exp\left\{-\frac{1}{2}\left(\frac{\eta^2}{\sigma_{s1}^2} + \frac{(\eta + V\xi)^2}{\sigma_{s2}^2}\right)\right\},$$  (11)

$$\psi_x(x, \xi, \eta_1) = \frac{1}{\sigma_{x1}(\xi,\eta) \cdot \sigma_{x2}(\xi,\eta)} \cdot exp\left\{-\left(\frac{1}{\sigma_{x1}^2(\xi,\eta)} + \frac{1}{\sigma_{x2}^2(\xi,\eta)}\right) \cdot \frac{x^2}{2}\right\}.$$  (12)

The function $\psi_y(y, \xi, \eta_1)$ is similar to $\psi_x(x, \xi, \eta_1)$ with replacement of the index and the arguments $x$ and $y$. An integration of $\psi_x$ over $dx$ yields:

$$\int_{-\infty}^{\infty} \psi_x(x, \xi, \eta_1) dx = \frac{\sqrt{2\pi}}{\sqrt{\sigma_{x1}^2(\xi,\eta) + \sigma_{x2}^2(\xi,\eta)}}.$$  (13)

An integral of $\psi_y$ over $dy$ is obtained with the replacement of $x$ by $y$.

Using the values of $\sigma_{x1,2}$ (4), we write

$$\sqrt{\sigma_{x1}^2(\xi,\eta) + \sigma_{x2}^2(\xi,\eta)} = \sqrt{\varepsilon_{x1}\beta_{x1}(\xi,\eta) + \varepsilon_{x2}\beta_{x2}(\xi,\eta)}.$$  (14)

The beta-functions $\beta_{xi}$ are defined in (5).

A similar expression for $\sigma_{y1,2}$ can be found by replacing $x$ by $y$. Substituting the values of the integral (13) and the denominator (14), as well as their $y$-analogs in (11), we arrive at a rather cumbersome expression for the luminosity of a collider with completely asymmetric bunches:

$$L = \frac{n_{bunch} N_1 N_2 f_0}{(2\pi)^2 \sigma_{s1} \sigma_{s2}} \cdot V \cdot \int_{-\infty}^{\infty} d\xi \int_{-\infty}^{\infty} d\eta \frac{exp\left\{-\frac{1}{2}\left(\frac{\eta^2}{\sigma_{s1}^2} + \left(\frac{\eta + V\xi}{\sigma_{s2}}\right)^2\right)\right\}}{\sqrt{\left(\varepsilon_{x1}\beta_{x1}(\xi,\eta) + \varepsilon_{x2}\beta_{x2}(\xi,\eta)\right)\left(\varepsilon_{y1}\beta_{y1}(\xi,\eta) + \varepsilon_{y2}\beta_{y2}(\xi,\eta)\right)}}.$$  (15)

*Note that beta-functions at particle collision points may have different values $(\beta_{x,y})_{1,2}$, if particles 1 and 2 differ in at least one of the parameters - charge, mass or energy. Since the values of the beta-functions are proportional to the magnetic rigidity of the particles, we can introduce the "relative magnetic rigidity parameter" $p$ of the colliding particles:*

$$p = \frac{\beta_1^*}{\beta_2^*} = \frac{p_1}{Z_1} \cdot \frac{Z_2}{p_2} \ . \tag{16}$$

Here $Z_{1,2}e$, $p_{1,2}$ are the charge and momentum of the particles 1 and 2. Parameter $p = 1$ in the case of different particles of the colliding beams if the focusing systems of both collider rings do not have in the common part of the rings (the interaction section) focusing elements defining values of the beta-functions.

Expression (15) is essentially simplified in the three particular cases considered below.

## 2. IDENTICAL COLLIDING BEAMS

In the case of identical beams we have:

$$v_1 = -v_2, \quad V = 2, \quad \beta_{x1}^* = \beta_{x2}^* \equiv \beta_x^*, \quad \beta_{y1}^* = \beta_{y2}^* \equiv \beta_y^*,$$
$$\sigma_{s1} = \sigma_{s2} \equiv \sigma_s, \quad \varepsilon_{x1} = \varepsilon_{x2} \equiv \varepsilon_x, \quad \varepsilon_{y1} = \varepsilon_{y2} \equiv \varepsilon_y. \tag{17}$$

After substitution in (15) the beta-functions (5) at accounting the expressions for $\xi$ and $\eta$ we find

$$L = \frac{n_{bunch} N_1 N_2 f_0}{8\pi^2 \sigma_s^2 \sqrt{\varepsilon_x \varepsilon_y \beta_x^* \beta_y^*}} \times \text{Int}_1(\beta_x, \beta_y, \sigma_s),$$

$$\text{Int}_1 = V \cdot \int_{-\infty}^{\infty} d\xi \int_{-\infty}^{\infty} d\eta \, \frac{1}{\sqrt{\left[1+\left(\frac{\xi+\eta}{\beta_x^*}\right)^2\right] \cdot \left[1+\left(\frac{\xi+\eta}{\beta_y^*}\right)^2\right]}} \cdot exp\left\{-\frac{\eta^2 + (\eta+2\xi)^2}{2\sigma_s^2}\right\}.$$

Now the velocities of colliding ions are equal by absolute value that gives us $V = 2$.

Transforming the numerator of the exponent:

$$\eta^2 + (\eta + 2\xi)^2 = 2[(\xi+\eta)^2 + \xi^2] \,,$$

passing from the integration over variables $\xi$, $\eta$ to the variables $\xi$ and $\varphi = \xi + \eta$, and calculating the Jacobian of the transition $\frac{D(\xi,\eta)}{D(\xi,\varphi)} = 1$ we find

$$\text{Int}_1 = 2 \cdot \int_{-\infty}^{\infty} e^{-\xi^2/\sigma_s^2} d\xi \int_{-\infty}^{\infty} \frac{d\phi}{\sqrt{\left[1+\left(\frac{\phi}{\beta_x^*}\right)^2\right] \cdot \left[1+\left(\frac{\phi}{\beta_y^*}\right)^2\right]}} \cdot e^{-\phi^2/\sigma_s^2} d\phi. \tag{18}$$

Thus, the double integral is transformed into the product of two independent integrals. The first of these, the integral over $\xi$, is equal to $\sqrt{\pi}$. In the case $\beta_x^* = \beta_y^* \equiv \beta^*$ we obtain from (18) the well-known expression for the luminosity of the collider for axially symmetric beams (do compare [1], formulas (6.134), (6.135)):

$$L = \frac{n_{bunch} N_1 N_2 f_0}{4\pi\sqrt{\varepsilon_x \varepsilon_y} \beta^*} \cdot F_{HG}, \quad F_{HG}(\alpha) = \frac{2}{\sqrt{\pi}} \int_0^{\infty} \frac{e^{-u^2} du}{1+(\alpha \cdot u)^2}, \quad \alpha = \frac{\sigma_s}{\beta^*}. \tag{19}$$

Here $F_{HG}(\alpha)$ is so called "Hour-glass function" describing luminosity dependence on $\sigma_s$ and $\beta^*$ parameters.

## 3. COLLISION OF THE COASTING BEAM WITH A BUNCH
## OF THE BUNCHED BEAM

As before, we consider the focusing system to be axially symmetric at the collision section, but the beams contain particles of different types and differ, generally speaking, in energy. We also assume:

$$\beta_{x1}^* = \beta_{y1}^* \equiv \beta_1^* \neq \beta_{x2}^* = \beta_{y2}^* \equiv \beta_2^*. \tag{20}$$

When one of the colliding beams ($N_2$) is coasting and has a uniform density along the circumference of the ring $C_{Ring}$ (see formula (2)) and another beam is bunched having $N_2$ particle in each of $n_{bunch}$ bunches, the formula for luminosity differs from (15) by a factor in the denominator of the fraction before the integral equal to $(2\pi)^{3/2}\sigma_s C_{Ring}$ now. Here $\sigma_s \equiv \sigma_{s1}$ is the Gaussian parameter of the bunch length of the bunched beam.

Confining ourselves to the case of axially symmetric beams we find that Formula (15) takes the form

$$L = \frac{n_{bunch}N_1 N_2 f_0}{(2\pi)^{3/2}\sigma_s C_{Ring}} \times \text{Int}_2(\beta_x, \beta_y, \sigma_s),$$

$$\text{Int}_2 = V \cdot \int_{-\infty}^{\infty} d\xi \int_{-\infty}^{\infty} d\eta \, \frac{1}{\varepsilon_1 \beta_1^*\left[1+\left(\frac{\xi+\eta}{\beta_1^*}\right)^2\right]+\varepsilon_2 \beta_2^*\left[1+\left(\frac{\xi+\eta}{\beta_2^*}\right)^2\right]} \cdot exp\left\{-\frac{\eta^2}{2\sigma_s^2}\right\}. \tag{21}$$

One should note that, in general, parameter $V$ (9) here can differ from 2.

We assumed here that the bunch length and the IP beta-functions are much smaller than the length of interaction region, i.e. we met the condition

$$\sigma_s, \; \beta_{1,2}^* << C_{Ring}. \tag{22}$$

Let's repeat the transformation of variables in the same way as above:

$$\varphi = \xi + \eta, \; \psi = \eta.$$

Hence $\xi = \psi - \varphi$, $\eta = \psi$ and the Jacobian of the transition $\frac{D(\xi,\eta)}{D(\varphi,\psi)} = 1$. As a result,

$$\text{Int}_2 = V \cdot \int_{-\infty}^{\infty} e^{-\frac{\psi^2}{2\sigma_s^2}} d\psi \int_{-\infty}^{\infty} \frac{d\phi}{\varepsilon_1 \beta_1^*\left[1+\left(\frac{\phi}{\beta_1^*}\right)^2\right]+\varepsilon_2 \beta_2^*\left[1+\left(\frac{\phi}{\beta_y^*}\right)^2\right]} \cdot \tag{23}$$

Integrating, we finally obtain

$$L = \frac{n_{bunch}N_1 N_2 f_0}{2 C_{Ring}} \cdot V \cdot \sqrt{\frac{\beta_1^* \beta_2^*}{(\varepsilon_1 \beta_1^* + \varepsilon_2 \beta_2^*)(\varepsilon_1 \beta_2^* + \varepsilon_2 \beta_1^*)}}. \tag{24}$$

The expression (24) can be simplified using the parameter $p$ (16):

$$L = \frac{n_{bunch} N_1 N_2 f_0}{2 C_{Ring}} \cdot V \cdot \sqrt{\frac{p}{(\varepsilon_1^2 + \varepsilon_2^2) + \varepsilon_1 \varepsilon_2 (1 + p^2)}} \ . \tag{25}$$

## 4. COLLISION OF TWO COASTING BEAMS

Under the same assumptions (22), we obtain from (15), similarly to (21), the luminosity value:

$$L = \frac{N_1 N_2 f_0}{2 C_1 C_2} \times \text{Int}_3(\beta_x, \beta_y, \sigma_s),$$

$$\text{Int}_3 = V \cdot \int_{-C_1/2}^{C_1/2} d\xi \int_{-\infty}^{\infty} d\eta \ \frac{1}{\varepsilon_1 \beta_1^* \left[1 + \left(\frac{\xi+\eta}{\beta_1^*}\right)^2\right] + \varepsilon_2 \beta_2^* \left[1 + \left(\frac{\xi+\eta}{\beta_2^*}\right)^2\right]} \cdot exp\left\{-\frac{\eta^2}{2\sigma_s^2}\right\}. \tag{26}$$

Here $C_1$ and $C_2$ are the circumferences of the storage rings, the collision frequency $f_0 = v_1/C_1$, regardless of the ratio between $v_1$ and $v_2$.

Integrating in the same way as in (23) - (25), we find

$$L = \frac{N_1 N_2 f_0}{2 C_2} \cdot V \cdot \sqrt{\frac{p}{p(\varepsilon_1^2 + \varepsilon_2^2) + \varepsilon_1 \varepsilon_2 (1 + p^2)}} \ . \tag{27}$$

## 5. SYNCHRONIZATION OF COLLISIONS AND THEIR FREQUENCY

Particle collisions occur in the same section common to both rings if the particle frequencies $f_{1,2}$ and number of the bunches ($n_{bunch1}$)$_{1,2}$ of two bunched beams satisfy equality

$$n_1 n_{bunch1} f_1 = n_2 n_{bunch2} f_2 \ , \tag{28}$$

where $n_{1,2}$ are integers. In the optimal case, the sizes of the rings and the energy of the particles are chosen so that $n_1 = n_2 = 1$. Under this condition the collision frequency is equal to

$$f_{coll} = n_{bunch1} f_1 = n_{bunch2} f_2. \tag{29}$$

This parameter should replace $n_{bunch} f_0$ in (11), (12), (15).

In the case of collision of two bunched beams the synchronization condition (28) is met at certain values of the colliding particles' energy (velocity $v_1, v_2$):

$$v_2 = v_1 \cdot \frac{n_1 n_{bunch1}}{n_2 n_{bunch2}} \cdot \frac{C_2}{C_1} \ . \tag{30}$$

However, the parameters $n_{1,2}$ are integer. Therefore, minimum variation of the particle energy is allowed with a step of $\Delta n = 1$. A scanning with a smaller step requires a special variation of the particle orbit in the ring. This is especially important for colliding beams of moderately relativistic particles, which occurs, for example, in electron-ion colliders [5].

If one of the colliding beams, either both, are coasting such a problem of synchronization does not exist.

If a coasting beam collides with a bunched one $f_{coll} = n_{bunch} f_0$, where $n_{bunch}$ and $f_0$ are parameters of the bunched beam, and when two coasting beams collide

$$f_{coll} = \min \{f_1, f_2\},$$

where $f_{1,2}$ are the revolution frequencies of the particles of the first and second beams.

We note that the obtained expressions for the luminosity (15), (24), (25), (27) are valid for both counter propagating and co-propagating beams. In the last case $V < 1$.

An important characteristic of luminosity is its distribution along the length of the interaction region. As can be seen from the integrands in the formulas for $Int_1$ (18) and $F_{HG}(\alpha)$ (19), $Int_2$ (21), (23) and $Int_3$ (26), the characteristic length $\Delta s$ of the function $dL/ds(s)$ is determined by minimal values of beta functions (if the length of the collision area is not limited intentionally):

$$\Delta s \approx \min\left\{\beta_1^*, \beta_2^*\right\} .$$

The author thanks Valery Lebedev, Janna Maltseva, Sergei Nagaitsev, and Peter Shatunov for very fruitful discussions and useful notes.